\documentclass{article}
\usepackage{ijcai21}

\usepackage{times}
\usepackage{soul}
\usepackage{xspace}
\usepackage{url}
\usepackage[hidelinks]{hyperref}
\usepackage[utf8]{inputenc}
\usepackage[small]{caption}
\usepackage{graphicx}
\usepackage{amsmath}
\usepackage{amsthm}
\usepackage{eucal}
\usepackage{amsfonts}
\usepackage{booktabs}
\usepackage{algorithm}
\usepackage{algorithmic}
\usepackage{multirow}
\usepackage{txfonts}
\usepackage{amsthm,amsmath,amssymb}
\usepackage{mathrsfs}

\title{Improving Sequential Recommendation Consistency with \\ Self-Supervised Imitation }

\author{
Xu Yuan$^{1,2,3}$ \footnote{Work done at JD.com.}\and
Hongshen Chen$^3$\and \\
Yonghao Song$^{1}$\and
Xiaofang Zhao$^{1}$\and 
Zhuoye Ding$^3$\and
Zhen He$^3$\And
Bo Long$^3$\\

\affiliations
$^1$Institute of Computing Technology, Chinese Academy of Sciences, Beijing, China\\
$^2$University of Chinese Academy of Sciences, Beijing, China\\
$^3$JD.com, China\\
\emails
\{yuanxu19g, songyonghao\}@ict.ac.cn,
ac@chenhongshen.com 
}

\begin{document}

\maketitle

\begin{abstract}
 Most sequential recommendation models capture the features of consecutive items in a user-item interaction history.
 Though effective, their representation expressiveness is still hindered by the sparse learning signals.
 As a result, the sequential recommender is prone to make inconsistent predictions.
 In this paper, we propose a model, \textbf{SSI}, to improve sequential recommendation consistency with $\textbf{\underline{S}}$elf-$\textbf{\underline{S}}$upervised $\textbf{\underline{I}}$mitation. 
 Precisely, we extract the consistency knowledge by utilizing three self-supervised pre-training tasks, where \textit{temporal consistency} and \textit{persona consistency} capture user-interaction dynamics in terms of the chronological order and persona sensitivities, respectively.
 Furthermore, to provide the model with a global perspective, \textit{global session consistency} is introduced by maximizing the mutual information among global and local  interaction sequences. 
 Finally, to comprehensively take advantage of all three independent aspects of consistency-enhanced knowledge, we establish an integrated imitation learning framework. 
 The consistency knowledge is effectively internalized and transferred to the student model by imitating the conventional prediction logit as well as the consistency-enhanced item representations.
 In addition, the flexible self-supervised imitation framework can also benefit other student recommenders.
 Experiments on four real-world datasets show that SSI effectively outperforms the state-of-the-art sequential recommendation methods.
\end{abstract}

\section{Introduction}



 With the development of mobile devices and Internet services, the past few years have seen the prosperity of a broad spectrum of recommendation systems (RS). It facilitates individuals making decisions over innumerable choices on the web.
 RS attracts a growing number of online retailers and e-commerce platforms to meet diverse user demands, enrich and promote their online shopping experiences.

 In real-world applications, users' current interests are affected by their historical behaviors. 
 When a user orders a smartphone, he/she will subsequently choose and purchase accessories like chargers, mobile phone covers, etc. 
 Such sequential user-item dependencies prevail and motivate the rising of sequential recommendation, where the user history interaction sequence is treated as a dynamic sequence. And the sequential dependencies are taken into account to characterize the current user preference to make more accurate recommendations\cite{10.1145/3159652.3159668}.

 As for the sequential recommendation, a surge of methods have been proposed to capture the sequential dynamics within user historical behaviors and predict the next use-interested item(s):
 Markov Chains\cite{10.5555/647235.720264}, recurrent neural networks (RNNs)\cite{hidasi2016sessionbased}, convolutional neural networks (CNNs)\cite{10.1145/3159652.3159656,10.1145/3289600.3290975}, Graph Neural Network(GNN)\cite{Wu_Tang_Zhu_Wang_Xie_Tan_2019}, self-attention mechanisms\cite{ijcai2018-546}.
 Markov chain-based models adopt K-order user-item interaction sequential transitions.
 CNN handles the transition within a sliding window, whereas RNN-based methods apply GRU or LSTM to compress dynamic user interests with hidden states.
 GNN-based methods take advantage of directed graphs to model complex user-item transitions in structured relation datasets. 
 And self-attention mechanisms emphasize relevant and essential interactions with different historical user-item interaction weights.

 Though previous methods show promising results, current sequential recommendation system learning relies heavily on the observed user-item action prediction. As a result, their representations and expressiveness are still limited.
 Such representation learning signal is too sparse to train expressive-enough item representations \cite{10.1145/1772690.1772773,10.1145/3357384.3357925}. 
 Not to mention that those subtle but diverse persona differences among users are also neglected.
 In this paper, we attempt to abstract the consistency knowledge with self-supervised pre-training tasks by exploring three aspects of consistency.
 To handle user-interaction dynamics, we propose to abstract consistency information with temporal and persona consistency, where 
 \textbf{temporal consistency} reflects that the recommender is expected to organize and display the items in a proper order to satisfy users' interests. 
 \textbf{Persona consistency} stands on the fact that the recommender should be capable of perceiving those diverse persona distinctions from the user-item interaction sequence. 

 Temporal and persona consistency knowledge is quite a straightforward approach to improve sequential recommendation systems by modeling instant interaction dynamics. However, long-term global session consistency is overlooked. 
 Without a global perspective, the model still suffers from noisy actions and makes inconsistent predictions.
 A typical case is that when a user unconsciously clicks wrong items, the system will be easily affected by short-term click and instantly make unrelated predictions.
 To mitigate such defect, we further bring \textbf{global session consistency} into the sequential recommendation, which enhances the item representations by maximizing the mutual information between global and local parts of the interaction sequence.
  
 Though self-supervised consistency knowledge can be quite beneficial, directly optimizing the consistency is intractable. 
 It's because training is performed on the whole session, which is not attainable in inference. 
 Furthermore, comprehensively integrating all three aspects of consistency pre-training knowledge plays a vital role in estimating user's preferences more accurately and consistently. 
 Therefore, we propose an integrated imitation learning framework. Consistency-enhanced pre-training models serve as teachers, and a sequential recommendation model is treated as a student.
 The consistency knowledge is effectively internalized and transferred to the student model by imitating the conventional prediction logit as well as the consistency-enhanced item representations.
 Additionally, with the flexibility of $\textbf{\underline{S}}$elf-$\textbf{\underline{S}}$upervised $\textbf{\underline{I}}$mitation (SSI) framework, the self-supervised learning can be easily transferred to other student recommenders as demand. 
 

 Experiments on four real-world datasets show that our self-supervised imitation framework effectively outperforms several state-of-the-art baselines. 

\section{Methodology}

\subsection{Problem Statement}
 Suppose $\mathcal{I}$ is the set of items and $\mathcal{U}$ is the set of users. For a specific user $u\in\mathcal{U}$, list $I_u=[q_1^{(u)},q_2^{(u)}...,q_k^{(u)}]$ denote the corresponding interaction sequence in chronological order where $q_k\in\mathcal{I}$ and $k$ is the time step. 
 Given the interaction history $I_u$, sequential recommendation aims to predict the item that user $u$ will interact with at time step $k + 1$.

\subsection{Teacher Base Model}
 Our teacher base model is a BERT-based structure\cite{10.1145/3357384.3357895}, built upon the popular self-attention layer. 
 The model randomly masks items in the user history interaction sequence in the training phase and replaces them with a unique token $\mathtt{[MASK]}$. Then predicts the original ids of the masked items based on the context. 
 In the test phase, the model appends the special token $\mathtt{[MASK]}$ at the end of the input sequence and then predicts the next item based on the final hidden representation of this token.

\subsection{Self-supervised Consistency Pre-training}

\begin{figure}[!t]
    \centering
    \includegraphics[scale=1.2]{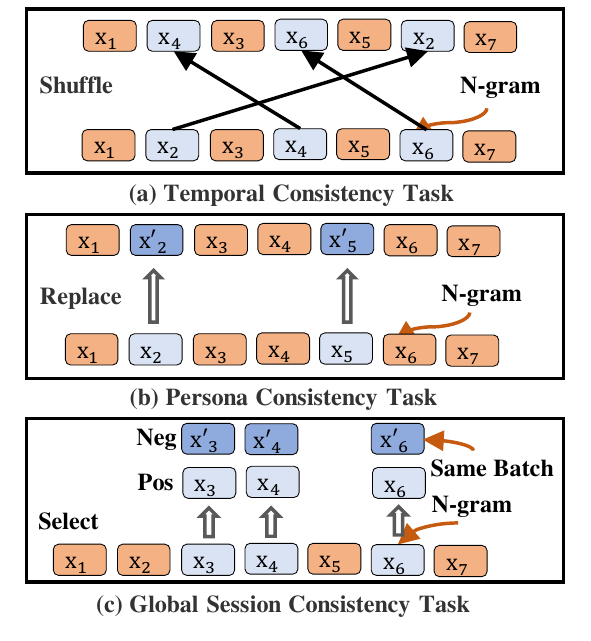}     
    \caption{Self-supervised consistency pre-training tasks. 
    (a) The order of some n-grams is shuffled in the sequence for the chronological prediction task.
    (b) Some n-grams in the sequence are replaced with other users' interaction n-grams. 
    (c) The mutual information is maximized between some n-grams and the rest of the sequence.
    }
    
    \label{fig:model}
\end{figure}

 Existing studies\cite{hidasi2016sessionbased,10.1145/2959100.2959167,8594844,logeswaran2018efficient} mainly emphasize the effect of sequential characteristics using user-item action prediction objective alone. 
 Solely relying on such sparse learning signals may result in under-learned item representations.
 We enrich learning representations' expressiveness by extracting consistency knowledge with self-supervised learning objectives to mitigate such defects.
 As exemplified in Figure \ref{fig:model}, we introduce three aspects of recommendation consistency: temporal, persona and global session consistency, respectively, where temporal consistency and persona consistency capture user-interaction dynamics and global session consistency enhances the model with the global perspective. 


\subsubsection{Temporal Consistency}
 Temporal consistency captures users' interaction sequence so that the recommendation system can display the items in a proper order that will satisfy users' interests better.
 Therefore, we design a user-item interaction sequence chronological order recognition task to enhance the model temporal sensitivity.
 First, we draw positive and negative samples by randomly exchanging the order of some n-gram sub-sequences in the user's interaction history with Bernoulli distribution. 
 Then we add a unique token named $\mathtt{[INT]}$ at the end of the sequence and let the corresponding hidden representation go through a classifier to predict whether the interaction sequence is in the original order or not. 
 The loss of temporal consistency is formulated as:
 \begin{equation}
 \mathcal{L}_{temp} = \sum_{i \in \mathscr{D}} - y_i^t \log{h^t(x_i^t)} - (1-y_i^t) \log(1-h^t(x_i^t)),
 \end{equation}
 where $\mathscr{D}$ is the set of input, $x_i^t$ is the output embedding, $y_i^t$ is label, $\mathcal{L}_{temp}$ is calculated by cross-entropy, and $h^t(\cdot)$ is a MLP.

\subsubsection{Persona Consistency}
 Person consistency models the diverse persona distinctions among different users.
 Positive samples are the interactions from one user. In contrast, for negative samples, we replace some user-item interaction n-grams with other items. 
 Similarly, we add a unique token $\mathtt{[INT]}$ at the end of the sequence, and a classifier predicts whether the user will interact with the current sequence.
 Persona consistency knowledge is perceived by differentiating whether the given user-item sequence is from one user.
 The loss of persona consistency is formulated as:
 \begin{equation}
 \mathcal{L}_{pers} = \sum_{i\in\mathscr{D}}-y_i^p\log{h^p(x_i^p)}-(1-y_i^p)\log(1-h^p(x_i^p)),
 \end{equation}
 where $\mathscr{D}$ is the set of input, $x_i^p$ is the output embedding, $y_i^p$ is label, $\mathcal{L}_{pers}$ is calculated by cross-entropy, and $h^p(\cdot)$ is a MLP.
  
\subsubsection{Global Session Consistency}
 Researches on image representation learning show that mutual information maximization between an image representation and local regions improves the quality of the representation\cite{dim}.
 Similar observations are also found in sentence representation learning \cite{kong2019mutual}.
 Inspired by previous studies, to avoid noise and provide the model with a global perspective, we further introduce the global session consistency as a self-supervision task for the sequential recommendation. 
 Given a user interaction sequence $I=[q_1,q_2...,q_k]$, we consider the local representation to be the encoded representations of n-grams in the sequence. The global representation is the rest of the masked sequence representation, which corresponds to the last token hidden state.
 We maximize the mutual information between the global representation and local representation.
 Denote an n-gram by $q_{n:m}$ and a masked sequence at position $n$ to $m$ by $\hat{q}_{n:m}$. We define $\mathcal{L}_{global}$ as:
\begin{equation}
\begin{aligned}
\mathcal{L}_{global} &=- \mathbb{E}_p(\hat{q}_{n:m},q_{n:m})[d(\hat{q}_{n:m})^{\top}d(q_{n:m}) \\
&- \log\sum_{\tilde{q}_{n:m} \in \tilde{S}}\exp (d(\hat{q}_{n:m})^{\top}d(\tilde{q}_{n:m})) ],
\end{aligned}
\end{equation}
 where $\tilde{q}_{n:m}$ is an n-gram from a set $\tilde{S}$ that consists of the positive sample $\hat{q}_{n:m}$ and negative n-grams from other sequence in the same batch, $d(\cdot)$ is the base model.

\subsubsection{Consistency Pre-Training}
 To ensure that each aspect of consistency is well-learned, we establish three independent models upon BERT4Rec\cite{10.1145/3357384.3357895} to induce temporal consistency, persona consistency, and global consistency information. 
The training objectives are defined as:
\begin{equation}
\begin{aligned}
\mathcal{L}_{I}^1 &= \mathcal{L}_{MLM} + \lambda_1\mathcal{L}_{temp}, \\
\mathcal{L}_{I}^2 &= \mathcal{L}_{MLM} + \lambda_2\mathcal{L}_{pers}, \\
\mathcal{L}_{I}^3 &= \mathcal{L}_{MLM} + \lambda_3\mathcal{L}_{global}, 
\end{aligned}
\end{equation}
 where $\mathcal{L}_{MLM}$ is the loss of the base model, $\mathcal{L}_{I}^i$ is the loss of each model. 
 $\mathcal{L}_{temp}$, $\mathcal{L}_{pers}$ and $\mathcal{L}_{global}$ are augmented as a kind of regularization to abstract the consistency knowledge from the sequence.
 $\lambda_i$ is a hyper-parameter balancing the importance of the two losses.

\begin{figure}[!thp]
    \centering
    \includegraphics[scale=0.35]{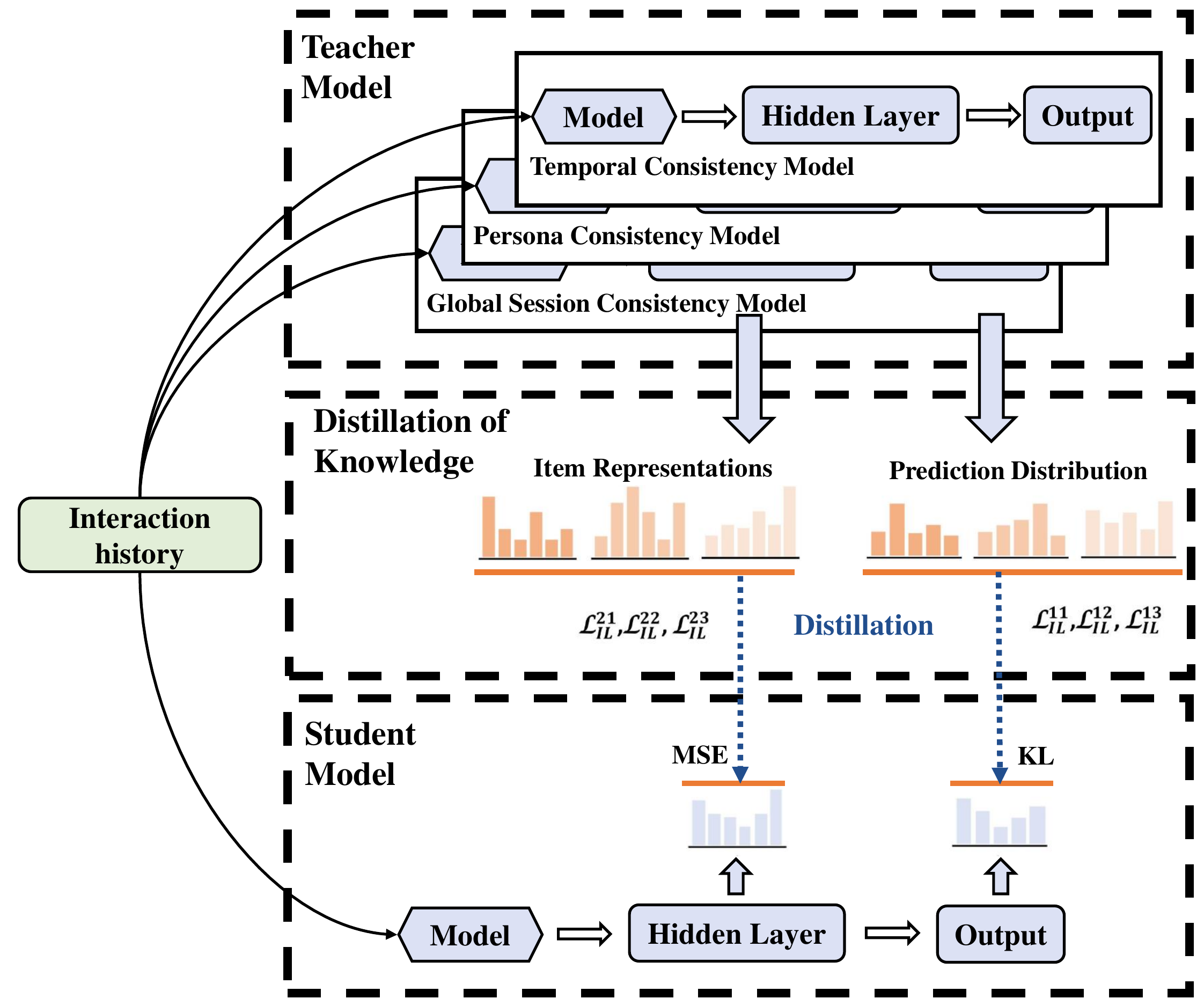}     
    \caption{Self-supervised imitation learning framework. Teacher model: three consistency-enhanced models. Student model: sequential recommendation model.}
    \label{fig:imitation}
\end{figure}

\subsection{Self-supervised Imitation Learning }
 So far, three aspects of consistency knowledge acquired from the proposed self-supervised learning tasks.
 However, sequential recommendation system still faces two major roadblocks in augmenting them: 1) in our approach, consistency learning is conducted on the entire user-item interaction session, whereas only the historical user behaviors are available during inference and future interactions are unattainable; 2) given three aspects of consistency-enhanced representations, it is paramount to explore a comprehensive integration of them to estimate a user's preference with better accuracy and
consistency.

 We propose an imitation learning framework to overcome the obstacles, where we treat the consistency-enhanced models as teachers and another sequential recommendation model as a student. The student is designated to distill consistency knowledge from three self-supervision-enhanced teachers by performing both the prediction distribution imitation and item representation imitation.
 Figure \ref{fig:imitation} illustrates the general self-supervised imitation learning framework.

\subsubsection{Prediction Distribution Imitation}
 In prediction distribution imitation, the student model distills the merits from the teacher model with prediction behaviors.
 In particular, we minimize the KL divergence of the predicted probability distribution between the teacher and the student:
 \begin{equation}
\mathcal{L}^1_{IL}(\theta_1, \theta_2) = \mathrm{D_{KL}}(p(q=Q|\theta_1)\|p(q=Q|\theta_2),
\label{i1}
\end{equation}
 where $Q$ is the set of candidate items, $\theta_1$ and $\theta_2$ are the parameters of the teacher and student model, respectively.

\subsubsection{Item Representation Imitation}
 Item representation imitation learns from the teacher by imitating the consistency-enhanced representations.
 In detail, we restrict the item representations discrepancy between the student and the teacher:
 \begin{equation}
 \begin{aligned}
 \mathcal{L}^2_{IL}(\theta_1, \theta_2) = f(g^t(\theta_1),g^s(\theta_2)),
 \end{aligned}
 \label{i2}
 \end{equation}
 where $g^t$ and $g^s$ are the outputs of item representations in teacher and student respectively, $f(\cdot)$ is the mean-squared-error (MSE) loss.

\subsubsection{Integrated Imitation Training}
 To effectively integrate three independent consistency teachers with the student, given the loss $\mathcal{L}_{IL}^1$ in Equation \ref{i1} and the loss $\mathcal{L}_{IL}^2$ in Equation \ref{i2}, the final objective function of the student is formulated as:
\begin{equation}
\begin{aligned}
\mathcal{L'} = \mathcal{L}_{s} + \sum_{i=1}^n\lambda_i(\mathcal{L}_{IL}^{1i}+\mathcal{L}_{IL}^{2i}),
\end{aligned}
\end{equation}
 where $\mathcal{L}_{s}$ is the loss of student model, $\lambda_i$ is the teacher importance weight, $n$ is the number of teachers.
 To further enable the model to control the effects of the teacher adaptively, we set $\lambda_i$ to be learnable parameters that are optimized during training regularization:
\begin{equation}
\begin{aligned}
\mathcal{L} = \mathcal{L'}+\sum_{i=1}^n\frac{1}{\lambda_i} + (\sum_{i=1}^n\frac{1}{\lambda_i})^2,
\end{aligned}
\end{equation}
 where the initial value of $\lambda_i$ is set to 1. 
 The $1/\lambda_i$ and $1/\lambda_i^2$ regularizations induce the model to learn to choose more effective teachers.
 
 Note that, with the flexibility of the self-supervised imitation framework, consistency knowledge can be easily transferred to many student recommenders.


\section{Experiments}


\subsection{Experimental Settings}

\subsubsection{Data Sets}
 We conduct our experiments on real-world datasets from Amazon review datasets.
 In this work, we select four subcategories: ``Beauty'', ``Toys and Games'', ``Sports and Outdoors'' and ``Musical Instruments''.  
 We select users with at least five interaction records for experiments.
 The data statistics after preprocessing are listed in Table \ref{Table:data}.

\begin{table}[!ht]
\centering
\resizebox{0.85\columnwidth}{!}{
\begin{tabular}{@{}lcccc@{}}
\toprule
Dataset & Beauty & Toys & Sports & Musical \\ \midrule
\# Users & 22,363 & 19,412 & 35,598 & 10,057 \\
\# Items & 12,102 & 11,925 & 18,358 & 36,098 \\
\# Actions & 176,139 & 148,185 & 260,739 & 77,733 \\
\# Avg. Actions / User & 7.9 & 7.6 & 7.3 & 7.3 \\
\# Avg. Actions / Item & 14.6 & 12.4 & 14.2 &  2.2 \\ \bottomrule
\end{tabular}}
\caption{The statistics of four datasets after preprocessing.}
\label{Table:data}
\end{table}

 To evaluate the recommendation's performance, we split each dataset into training/validation/testing sets. 
 We hold out the last two interactions as validation and test sets for each user, while the other interactions are used for training.

\subsubsection{Evaluation Metrics}
 Usually, the recommendation system suggests a few items at once. The desired item should be amongst the first few listed items. 
 Therefore, we employ Recall@N and NDCG@N \cite{ndcg} to evaluate the recommendation performance.
 In general, higher metric values indicate better ranking accuracy. 
 Recall@k indicates the proportion of cases when the rated item is amongst the top-k items.
 NDCG@k is the normalized discounted cumulative gain at k, which takes the rank of recommended items into account and assigns larger weights on higher positions.
 To avoid high computation cost on all user-items in evaluation, following the strategy in \cite{ijcai2019-190}, we randomly draw 99 negative items that have not been engaged with the user and rank the ground-truth item among the 100 items.

\begin{table*}[!ht]
\centering
\resizebox{1.56\columnwidth}{!}{
\begin{tabular}{@{}clcccccccc@{}}
\toprule
Dataset & Baseline & {Recall@5} & {NDCG@5} & {Recall@10} & NDCG@10 & {Recall@15} & {NDCG@15} & {Recall@20} & {NDCG@20} \\ \midrule
 & MostPop & 0.204 & 0.124 & 0.297 & 0.172 & 0.346 & 0.188 & 0.412 & 0.197 \\
 & S$^{3}$-Rec & 0.369 & 0.285 & 0.470 & \underline{0.321} & 0.539 & \underline{0.338} & 0.588 & \underline{0.351} \\
 & GREC & 0.375 & 0.282 & 0.470 & 0.314 & 0.533 & 0.333 & 0.576 & 0.343 \\
 & BERT4Rec & \underline{0.378} & \underline{0.290} & \underline{0.471} & 0.320 & \underline{0.541} & \underline{0.338} & \underline{0.591} & 0.350 \\
 & GRU4Rec & 0.279 & 0.206 & 0.366 & 0.234 & 0.429 & 0.250 & 0.482 & 0.263 \\
 & HGN & \underline{0.378} & 0.288 & 0.470 & 0.317 & 0.531 & 0.334 & 0.580 & 0.345 \\
 & $SSI_{GRU4Rec}$ & 0.385 & 0.292 & 0.473 & 0.326 & 0.548 & 0.343 & 0.604 & 0.356 \\
\multirow{-8}{*}{Beauty} & SSI & \textbf{0.389} & \textbf{0.298} & \textbf{0.487} & \textbf{0.329} & \textbf{0.551} & \textbf{0.345} & \textbf{0.605} & \textbf{0.358} \\ \midrule
 & MostPop & 0.182 & 0.103 & 0.235 & 0.141 & 0.299 & 0.169 & 0.332 & 0.178 \\
 & S$^{3}$-Rec & 0.350 & 0.265 & 0.458 & 0.303 & \underline{0.530} &\underline{0.332} & \underline{0.586} & \underline{0.336} \\
 & GREC & 0.359 & 0.262 & 0.460 & 0.295 & 0.525 & 0.319 & 0.576 & 0.329 \\
 & BERT4Rec & 0.353 & 0.268 & 0.459 & 0.302 & 0.529 & 0.321 & 0.584 & 0.334 \\
 & GRU4Rec & 0.216 & 0.146 & 0.313 & 0.177 & 0.381 & 0.195 & 0.438 & 0.209 \\
 & HGN & \underline{0.364} & \underline{0.275} & \underline{0.463} & \underline{0.307} & 0.528 & \underline{0.325} & 0.580 & 0.335 \\
 & $SSI_{GRU4Rec}$ & 0.366 & 0.270 & 0.465 & 0.308 & 0.532 & 0.328 & 0.588 & 0.342 \\
\multirow{-8}{*}{Toys} & SSI & \textbf{0.369} & \textbf{0.280} & \textbf{0.470} & \textbf{0.313} & \textbf{0.539} & \textbf{0.331} & \textbf{0.595} & \textbf{0.344} \\ \midrule
 & MostPop & 0.199 & 0.122 & 0.274 & 0.145 & 0.301 & 0.183 & 0.368 & 0.194 \\
 & S$^{3}$-Rec & 0.345 & \underline{0.252} & 0.462 & 0.285 & 0.540 & 0.307 & \underline{0.605} & 0.324 \\
 & GREC & 0.288 & 0.203 & 0.378 & 0.241 & 0.464 & 0.267 & 0.502 & 0.282 \\
 & BERT4Rec & \underline{0.347} & \underline{0.252} & \underline{0.463} & \underline{0.290} & \underline{0.541} & \underline{0.311} & 0.603 & \underline{0.325} \\
 & GRU4Rec & 0.234 & 0.160 & 0.335 & 0.193 & 0.405 & 0.211 & 0.464 & 0.225 \\
 & HGN & 0.309 & 0.227 & 0.416 & 0.261 & 0.489 & 0.281 & 0.547 & 0.294 \\
 & $SSI_{GRU4Rec}$ & 0.357 & 0.254 & 0.467 & 0.296 & 0.549 & 0.318 & 0.608 & 0.329 \\
\multirow{-8}{*}{Sports} & SSI & \textbf{0.361} & \textbf{0.263} & \textbf{0.479} & \textbf{0.301} & \textbf{0.558} & \textbf{0.322} & \textbf{0.619} & \textbf{0.336} \\ \midrule
 & MostPop & 0.126 & 0.048 & 0.179 & 0.079 & 0.206 & 0.103 & 0.236 & 0.117 \\
 & S$^{3}$-Rec & 0.243 & \underline{0.179} & 0.322 & \underline{0.206} & 0.374 & 0.211 & 0.419 & 0.224 \\
 & GREC & 0.232 & 0.175 & 0.298 & 0.194 & 0.356 & 0.211 & 0.401 & 0.223 \\
 & BERT4Rec & \underline{0.244} & 0.177 & \underline{0.326} & 0.204 & \underline{0.381} & \underline{0.218} & \underline{0.422} & \underline{0.228} \\
 & GRU4Rec & 0.199 & 0.148 & 0.279 & 0.174 & 0.341 & 0.190 & 0.394 & 0.203 \\
 & HGN & 0.227 & 0.171 & 0.299 & 0.194 & 0.355 & 0.209 & 0.399 & 0.219 \\
 & $SSI_{GRU4Rec}$ & 0.234 & 0.172 & 0.328 & 0.202 & 0.389 & 0.218 & 0.428 & 0.228 \\
\multirow{-8}{*}{Musical} & SSI & \textbf{0.246} & \textbf{0.188} & \textbf{0.331} & \textbf{0.213} & \textbf{0.388} & \textbf{0.228} & \textbf{0.427} & \textbf{0.239} \\ \bottomrule
\end{tabular}
}
\caption{Performance comparison of baselines and our approaches, where our approach SSI's best results are in bold. The underlined numbers are the best results besides SSI.}  
\label{table2}  
\end{table*}


 \begin{figure*}[!ht]
    \centering
    \includegraphics[scale=0.17]{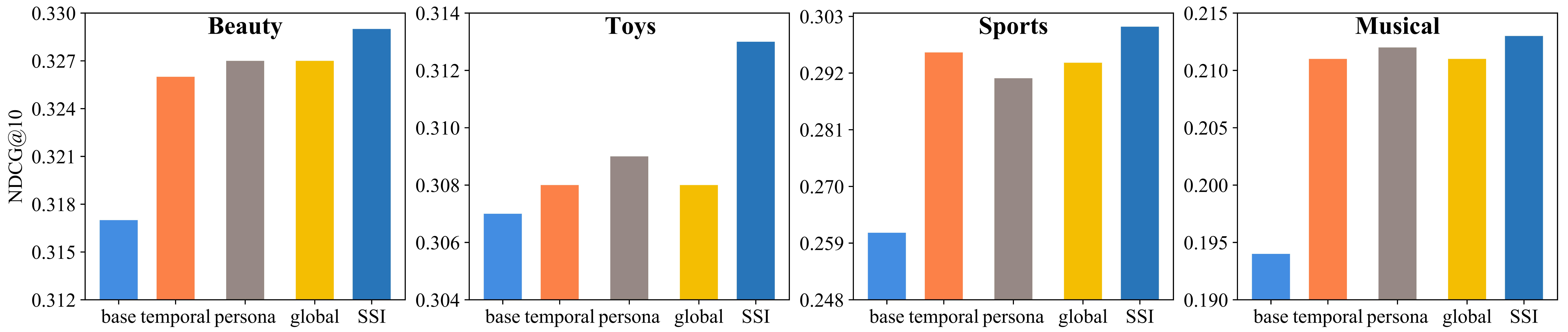}     
    \caption{Our best model's performance comparison with different aspects of consistency knowledge on four datasets (NDCG@10).}
    \label{fig:result2}
\end{figure*}


\subsubsection{Baselines}

 We compare with the following sequential recommendation models to justify the effectiveness of our approaches.

\begin{itemize}
\item \textbf{MostPop} recommends the item according to the popularity measured by the number of user-interactions, which provides a simple non-personalized baseline. 

\item \textbf{GRU4Rec}\cite{hidasi2016sessionbased} adopts GRU to capture sequential dependencies and makes predictions for session-based recommendation.

\item \textbf{BERT4Rec}\cite{10.1145/3357384.3357895} learns a bidirectional representation model to make recommendations by learning a cloze objective with reference to BERT.

\item \textbf{HGN}\cite{10.1145/3292500.3330984} captures both the long-term and short-term user interests by hierarchical gating network.

\item \textbf{S$^{3}$-Rec}\cite{10.1145/3340531.3411954} is a recently proposed self-supervised learning framework under mutual information maximization principle for the sequential recommendation, which is based on self-attentive neural architecture.

\item \textbf{GREC}\cite{yuan2020future} is an encoder-decoder framework that trains the encoder and decoder by a gap-filling mechanism.
\end{itemize}

\subsubsection{Parameter Settings}
 For MostPop and GRU4Rec, we implement them with PyTorch. 
 For S$^{3}$-Rec, we use two auxiliary self-supervised objectives without extra attribute information to ensure fairness. 
 For other methods, we use the source code provided by their authors. 
 All hyper-parameters are set following the suggestions from the original papers. 
 For SSI, our teacher model bases on BERT4Rec\cite{10.1145/3357384.3357895} and our student model bases on HGN\cite{10.1145/3292500.3330984}. 
 We set the number of the self-attention blocks and the attention heads as 8 and 4.
 The dimension of the embedding is 256. 
 The hyper-parameters are set as $\lambda_1 = \lambda_2 = \lambda_3 = 1$.
 We use the Adam optimizer\cite{adam} with a learning rate of 0.001, where the batch size is set as 256 in the teacher and student model, respectively.

\subsection{Results and Analysis}
\subsubsection{Overall Results}
 Table \ref{table2} presents the performance comparisons between several baselines and our model (SSI). 
 SSI consistently achieves the best performance on four datasets with all evaluation metrics, verifying our model's superiority.
 Compared with BERT4Rec, our model shows much better performance. In our model, bidirectional item representations are further improved with three aspects of consistency. The integrated imitation network effectively distills consistency-enhanced pre-training knowledge into the student model, which combines the merits of the pre-training teachers and student network. 
 SSI outperforms HGN. The reason is that HGN undertakes the student network in our model. The performance of HGN is effectively improved by imitating self-supervised consistency knowledge.
 The performance of S$^{3}$-Rec is inferior to SSI. Although the absence of extra attribute information is a potential influencing factor, another primary reason is that S$^{3}$-Rec ignores the consistency of sequential recommendation.

\subsubsection{Flexibility of Self-supervised Imitation}
 In our self-supervised imitation framework, the student recommender can be any sequential recommendation system.
 To verify its flexibility, we also list the performance of GRU4Rec as the student model.
 In Table \ref{table2}, we notice that GRU4Rec and HGN are both improved by self-supervised imitation (SSI$_{GRU4Rec}$, SSI), which demonstrates its applicability.
 More encouragingly, concerning the performance of the base model, GRU4Rec is much inferior to HGN. However, after being augmented with self-supervised imitation, the performance of SSI$_{GRU4Rec}$ is on par with SSI.
 In conclusion, the self-supervised consistency knowledge can be easily transferred to other student recommenders with the flexibility of the self-supervised imitation framework.

\begin{figure}[!ht]
    \centering
    \includegraphics[scale=0.12]{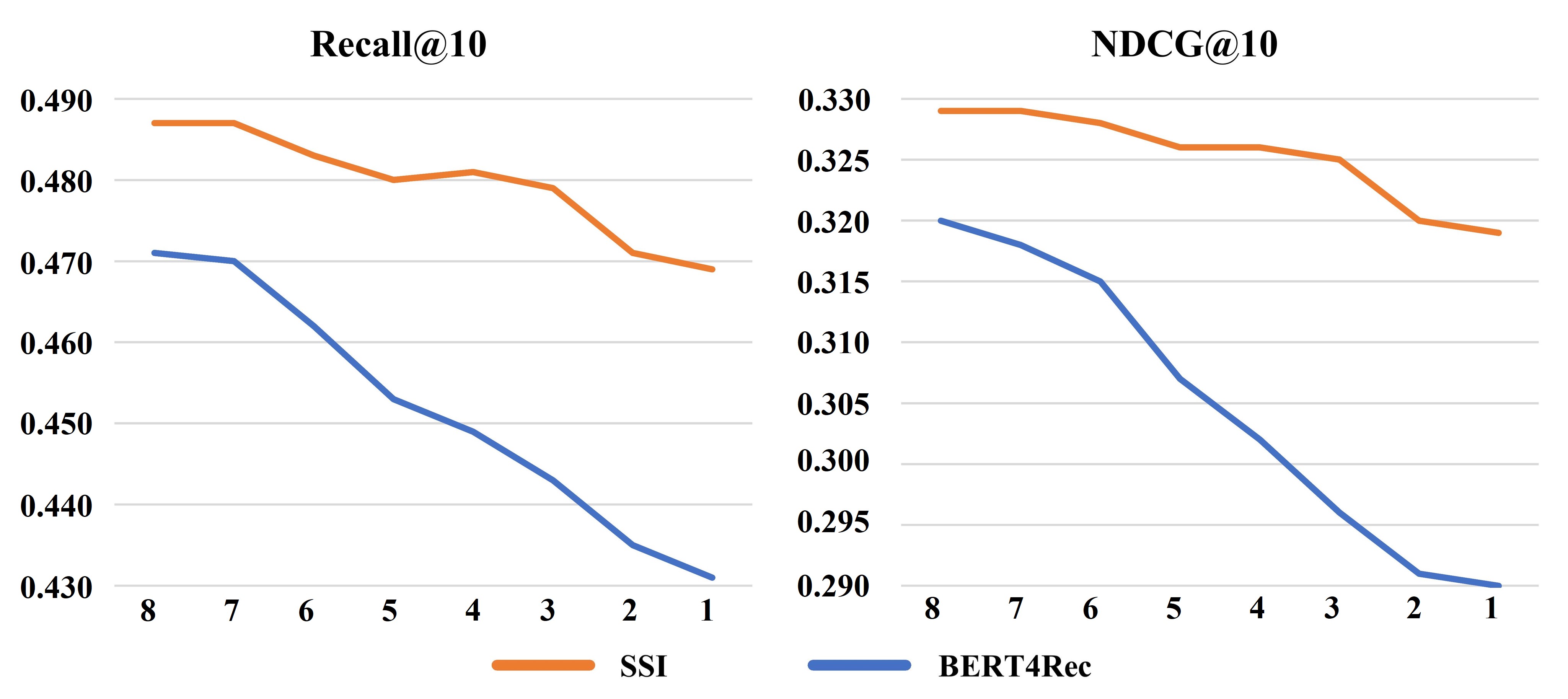}     
    \caption{Performance comparison between different number of self-attention layers on Beauty.}
    \label{fig:result3}
\end{figure}

\subsubsection{Impact of Different Aspects of Consistency}
 In this paper, we propose three aspects of self-supervised pre-training tasks to abstract consistency knowledge.
 Figure \ref{fig:result2} shows the comparison of effects between the proposed three aspects of consistency, temporal, persona, and global session consistency. 
 Compared with the base model, we observe that independently utilizing temporal, personal, or global session consistency effectively boots the model performance. The gains from different aspects of consistency vary on different datasets.
 However, jointly imitating all three aspects of consistency-enhanced knowledge achieves the best performance.

\subsubsection{Impact of Pre-training Model Size}
 In our model, consistency pre-training is performed upon BERT4Rec. 
 As self-attention based pre-training is computation-expensive, we would like to verify the model performance on different pre-training model layers.
 Figure \ref{fig:result3} illustrates the model performance comparison between our model and BERT4Rec along with different self-attention layers.
 We observe the performance degradation of both models along with the decreased number of self-attention layers. 
 What's more, the performance gap increases when comparing a relative deep model with eight self-attention layers and a shallow one with only one self-attention layer.
 Obviously, our model is more robust to computational cost than BERT4Rec. 
 We conjecture that this is because our model effectively enriches the item representation expressiveness with three consistency-distillation objectives.
 

\begin{table}[!ht]
\resizebox{1\columnwidth}{!}{
\begin{tabular}{@{}ccccc@{}}
\toprule
 & {Recall@10} & Improve & {NDCG@10} & Improve \\ \midrule
SSI & 0.487 &  & 0.329 &  \\
w/o item representation imitation & 0.483 & -0.8\% & 0.325 & -1.2\% \\
w/o prediction distribution imitation & 0.478 & -1.8\% & 0.321 & -2.4\% \\
w/o both & 0.470 & -3.5\% & 0.317 & -3.6\% \\ \bottomrule
\end{tabular}}
\caption{Effect of different imitation learning methods on Beauty.}  
\label{table:table3}  
\end{table}

\subsubsection{Impact of Integrated Imitation Learning}
 Integrated imitation learning transfers consistency-enhanced knowledge from three aspects of teachers to the student. 
 In our framework, we employ the prediction logit imitation as well as the item representation imitation. 
 Table \ref{table:table3} shows the performance differences. 
 Both kinds of imitations contribute to the final performance. 
 Though prediction distribution regularization is quite useful for consistency knowledge transferring, item representation imitation further improves the knowledge distillation efficiency.
 We speculate that item representation provides a short-cut for consistency knowledge distillation.

\section{Related Work}

 Nowadays, many approaches have been proposed to model the user's historical interaction sequence. 
 The methods based on the Markov chain predict the subsequent user interaction by estimating the probability of transfer matrix between items\cite{10.5555/647235.720264}.
 RNN-based methods model the sequential dependencies over the given interactions from left to right and make recommendations based on this hidden representation. 
 Except for the basic RNN, long short-term memory (LSTM)\cite{10.1145/3018661.3018689}, gated recurrent unit (GRU)\cite{hidasi2016sessionbased}, hierarchical RNN\cite{10.1145/3109859.3109896} have also been developed to capture the long-term or more complex dependencies in a sequence.
 CNN-based methods first embed this historical interactive information into a matrix and then use CNN to treat the matrix as an image to learn its local features for subsequent recommendation\cite{10.1145/3159652.3159656,10.1145/3289600.3290975}. 
 GNN-based methods first build a directed graph on the interaction sequence, then learn the embeddings of users or items on the graph to get more complex relations over the whole graph \cite{Wu_Tang_Zhu_Wang_Xie_Tan_2019}. 
 Attention models emphasize those important interactions in a sequence while downplaying those that have nothing to do with user interest\cite{ijcai2018-546}.

 Self-supervised learning\cite{albert,bert} prevails in language modelling. 
 It allows us to mine knowledge from unlabeled data in a supervised manner.
 S$^{3}$-Rec\cite{10.1145/3340531.3411954} enhances data representations and learns the correlations with mutual information maximization for the sequential recommendation, whereas we enrich the item representation expressiveness with the temporal, person and global session consistency, and distills the consistency enhanced knowledge to the student by imitation learning.

 Knowledge distillation\cite{hinton2015distilling} introduces soft-target related to the teacher network as a part of the loss to induce student network training and realize knowledge transfer. 
 Model compression\cite{10.1145/1150402.1150464} is a great application of knowledge distillation, which uses a light model to learn the knowledge of the heavy model to improve the efficiency of inference.
 In our work, we combine the benefits of self-supervised learning and imitation learning for the sequential recommendation, where three elaborately designed self-supervised consistency learning tasks transfer knowledge through integrated imitation learning.

\section{Conclusion}

 In this work, we improve sequential recommendation consistency with self-supervised imitation.
 First, three aspects of consistency knowledge are extracted with self-supervision tasks, where temporal and persona consistencies capture user-item dynamics, and the global session consistency provides a global perspective with interaction mutual information.
 Then an imitation framework integrates the consistency knowledge and transfers it to the student. 
 Due to its special merits of flexibility, consistency knowledge can easily benefit other student recommenders as demand.
 Experimental results and analysis demonstrate the superiority of the proposed model.
 

\section*{Acknowledgments}
This work was supported by National key R\&D Program of China (Grant No. 2018YFB0904503).  
Hongshen Chen and Yonghao Song are the corresponding authors.

\bibliographystyle{named}


\end{document}